\documentclass[aps,pra,showpacs,twocolumn,amsmath,amssymb]{revtex4}
\usepackage{graphicx}
\usepackage{amsmath}
\usepackage{color}
\usepackage{dcolumn}
\usepackage{bm}
\usepackage{booktabs}

\newcommand{\Xstate}{\mbox{$X^1\Sigma^+_g$}}
\newcommand{\astate}{\mbox{$a^3\Sigma^+_u$}}
\newcommand{\wn}{cm$^{-1}$}

\begin{document}

\title{Feshbach spectroscopy and analysis of the interaction potentials of ultracold sodium}

\author{S.\,Knoop}\email[s.knoop@vu.nl]{}\email[Present address: LaserLaB Vrije Universiteit, De Boelelaan 1081, 1081 HV Amsterdam, The Netherlands]{}
\author{T.\,Schuster}
\author{R.\,Scelle}
\author{A.\,Trautmann}
\author{J.\,Appmeier}
\author{M.\,K.\,Oberthaler}\affiliation{Kirchhoff-Institut f\"{u}r Physik, Universit\"{a}t Heidelberg, Im Neuenheimer Feld 227, 69120 Heidelberg, Germany}
\author{E.\,Tiesinga}\affiliation{Joint Quantum Institute, National Institute of Standards and Technology and University of Maryland, 100 Bureau Drive, Stop 8423 Gaithersburg, Maryland 20899-8423, USA}
\author{E.\,Tiemann}\email[tiemann@iqo.uni-hannover.de]{}\affiliation{Institute of Quantum Optics, Leibniz Universit\"{a}t Hannover, 30167 Hannover, Germany}

\date{\today}

\begin{abstract}
We have studied magnetic Feshbach resonances in an ultracold sample of Na prepared in the absolute hyperfine ground state. We report on the observation of three $s$-, eight $d$-, and three $g$-wave Feshbach resonances, including a more precise determination of two known $s$-wave resonances, and one $s$-wave resonance at a magnetic field exceeding 200\,mT. Using a coupled-channels calculation we have improved the sodium ground-state potentials by taking into account these new experimental data, and derived values for the scattering lengths. In addition, a description of the molecular states leading to the Feshbach resonances in terms of the asymptotic-bound-state model is presented.
\end{abstract}


\maketitle

\section{Introduction\label{Introduction}}

The observation of Feshbach resonances in ultracold atomic gases \cite{inouye1998oof,courteille1998ooa,roberts1998rmf,vuletic1999ool}
has opened a wealth of exciting experiments, as their presence allows tunability of the two-body interaction strength and coherent association of ultracold molecules.
Feshbach resonances are caused by the coupling between an atom pair and a molecular state in a closed channel potential \cite{chin2010fri}. The most commonly used ones are magnetically induced resonances that originate from a difference in the magnetic moment of the molecular state and the atomic threshold \cite{tiesinga1993tar}. The location of the resonances is very sensitive to the interaction potentials. Much effort has gone into constructing accurate two-body potentials in order to predict Feshbach resonances and scattering properties. Conversely, precise determination of Feshbach resonances can be regarded as a sensitive spectroscopic tool to map out the molecular spectrum of the least bound states.

For Na only three Feshbach resonances have been reported so far, namely two for
an ensemble prepared in the absolute hyperfine ground state $|F,m_F\rangle=|1,1\rangle$ \cite{inouye1998oof} and one in the hyperfine substate $|1,-1\rangle$ \cite{stenger1999sei}. All three are $s$-wave resonances, i.\,e.\ caused by molecular states with rotational quantum number $l=0$. In parallel, Raman spectroscopy on a molecular beam of sodium dimers to locate the least-bound rovibrational states of the ground state potentials \Xstate~and \astate~has been performed \cite{elbs1999oot,samuelis2000cac,laue2002mfi}. Scattering lengths \cite{samuelis2000cac} as well as Feshbach resonance positions \cite{laue2002mfi} were extracted. In addition, Refs.~\cite{fatemi2002ugs,araujo2003tcp} presented an extensive study of the lowest triplet potential \astate~by two-color photoassociation spectroscopy on magneto-optically trapped Na atoms. However, more precise Feshbach spectroscopy of Na is needed to improve the knowledge of the interaction properties.

Here, we report on the observation of seven $d$-wave and three $g$-wave Feshbach resonances (i.\,e.\ caused by molecular states with $l=2$ and $l=4$, respectively) in the magnetic field range between 49\,mT and 91\,mT, and a more precise determination of two $s$-wave Feshbach resonance positions first observed in Ref.\,\cite{inouye1998oof}. In addition, we have observed an $s$- and $d$-wave Feshbach resonance at 205.4\,mT and 159.0\,mT, respectively. Using coupled-channels calculations we have improved the Na$_2$ ground-states potentials by combining our new Feshbach resonance data with the known spectroscopy data. We have also applied the recently developed asymptotic-bound-state model (ABM) \cite{wille2008eau,tiecke2010abs} for the relevant $l=0$, 2 and 4 molecular states of Na$_2$, providing a simple description of the data. Note that $g$-wave Feshbach resonances have so far only been observed for Cs \cite{herbig2003poa,chin2004pfs}, for which even molecular states with $l=8$ have been populated \cite{mark2007sou,knoop2008mfm}.

\begin{figure*}
\includegraphics[width=18cm]{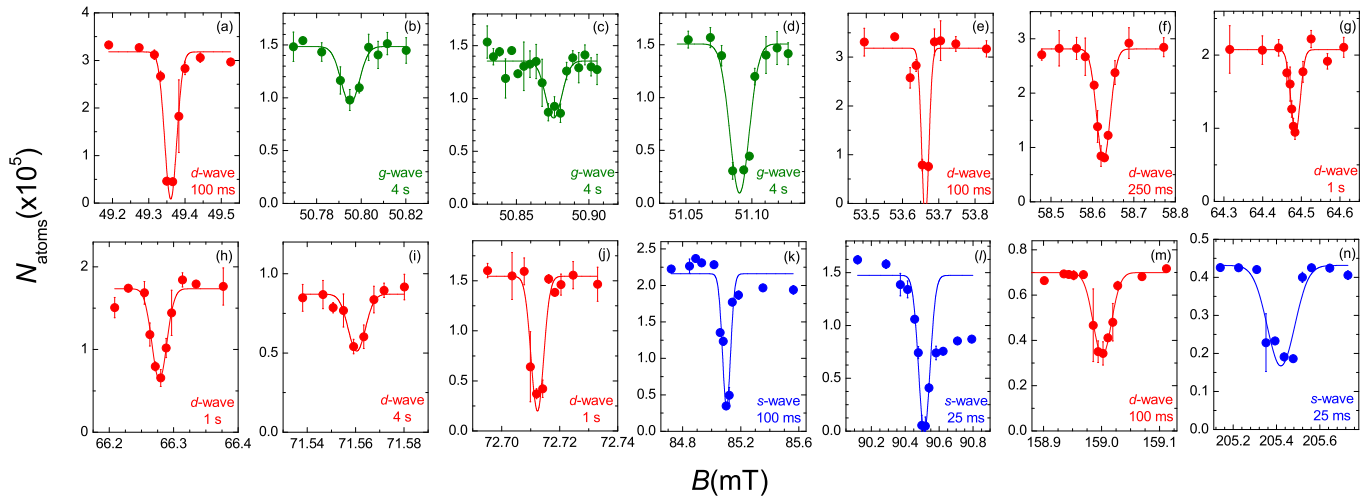}
\caption{(color online) Feshbach spectroscopy measurements for Na prepared in the lowest hyperfine substate $|F,m_F\rangle=|1,1\rangle$, showing the number of atoms observed after a certain hold time as function of the magnetic field $B$. The field range for each panel differs and longer hold times indicate weaker and narrower resonances. Blue, red and green colors corresponds to $s$-, $d$-, and $g$-wave resonances, respectively. Each data point is an average over 3 to 4 experimental runs; the error bars represent the one standard deviation statistical uncertainty. The solid lines are Gaussian fits, used to determine the loss resonance position. Note that for panel (\textit{l}) the right side of the resonance is not taken into account in the fit.
\label{FRall}}
\end{figure*}

This paper is organized as follows. In Sec.\,\ref{Exp} we outline our experimental procedure for the preparation of a Na Bose-Einstein condensate (Sec.\,\ref{Prep}) and the Feshbach spectroscopy measurements (Sec.\,\ref{FS}). In Sec.\,\ref{Theory} the theoretical model for the near threshold molecular levels of Na$_2$ is presented, introducing the interaction Hamiltonian (Sec.\,\ref{intro hamiltonian}), followed by a short description of the ABM and its application to the Feshbach resonance data (Sec.\,\ref{ABM}). We present results of the coupled-channel calculation deriving improved ground-state potentials of Na$_2$ (Sec.\,\ref{CC}). We summarize, give the scattering lengths obtained from the new potentials, and conclude in Sec.\,\ref{concl}.

\section{Experiment}\label{Exp}

\subsection{Preparation of a Na Bose-Einstein condensate}\label{Prep}

The observation of magnetically induced Feshbach resonances requires the preparation of a sufficiently high density and sufficiently cold atomic sample, on which an external homogeneous magnetic field can be applied. We use a Na Bose-Einstein condensate (BEC) in an optical dipole trap (ODT). Our experimental procedure is based on evaporative cooling of Na in the $|2,2\rangle$ state in a cylindrical symmetric cloverleaf Ioffe-Pritchard magnetic trap \cite{hadzibabic2003fii}.

We load about $10^9$ atoms from a dark-spot magneto-optical trap into the magnetic trap, and then spin-polarize at a bias offset field of 2.0\,mT into the $|2,2\rangle$ state, in analogy to the strategy of Ref.\,\cite{stam2006spc}. Residual atoms in $|2,1\rangle$ are removed state-selectively by microwave radiation, using an offset field of 5.3\,mT. Then, we apply forced evaporative cooling driving the $|2,2\rangle$ to $|1,1\rangle$ transition for about 30\,s. We first use a compressed magnetic trap with axial and radial trap frequencies $(\nu_{\rm ax}, \nu_{\rm rad}) = (33, 400) $\,Hz, while during the last few seconds of the evaporation stage we expand the magnetic trap to trap frequencies $(\nu_{\rm ax}, \nu_{\rm rad}) = (33, 130) $\,Hz in order to reduce losses from three-body recombination. After evaporation, we are left with $1\times10^7$ atoms at a temperature of 2\,$\mu$K.

At this point we load the atoms into an ODT, formed by two crossed laser beams with waists of 60\,$\mu$m each and powers of 1\,W and 0.5\,W, respectively, and derived from a 1064\,nm Nd:YAG laser. The resulting trap has a depth of about 8\,$\mu$K and frequencies  $(\nu_{\rm ax}, \nu_{\rm rad}) = (130, 310) $\,Hz. By our loading procedure and subsequent rethermalization by elastic collisions, we routinely obtain a BEC of $5\times10^5$ atoms in $|2,2\rangle$. Subsequently, we transfer all atoms to the state $|1,1\rangle$ by rapid adiabatic passage, applying a 50\,ms magnetic field sweep of 24\,$\mu$T around an offset field of 0.1\,mT at a fixed microwave frequency. The typical lifetime of the $|1,1\rangle$ BEC with peak density of $3 \times 10^{14}$\,cm$^{-3}$ in the ODT is about 5\,s, limited by three-body recombination.

\subsection{Feshbach spectroscopy}\label{FS}

A common way to probe Feshbach resonances is to measure the rate of atom loss as a function of magnetic field $B$. The dynamics in a Bose gas is controlled by collisions that are characterized by the $s$-wave scattering length $a$. Near a Feshbach resonance $a$ can be described as \cite{moerdijk1995riu}:
\begin{equation}
\label{scatteringlength}
a(B)=a_{\rm bg}\left(1-\frac{\Delta}{B-B_0}\right),
\end{equation}
where $a_{\rm bg}$ is the background scattering length, $B_0$ the resonance field and $\Delta$ the magnetic width of the resonance. The presence of a Feshbach resonance leads to a strong enhancement of loss because the dominant loss process, namely three-body recombination, scales strongly with $a$ \cite{fedichev1996tbr,esry1999rot,nielsen1999ler,braaten2001tbr,petrov2004tbp}.

To locate a resonance, we ramp to a magnetic field $B$, wait for a certain hold time, after which we ramp the field back to zero and use absorption imaging to measure the remaining number of atoms. The ramp times are always short compared to the hold times. A compilation of our measurements is shown in Fig.\,\ref{FRall} and the experimental results are summarized in Table\,\ref{FRlist}. We have observed 12 loss resonances between 49\,mT and 91\,mT, which we assign as seven $d$-wave and three $g$-wave resonances. The $s$-wave resonances at 85.1\,mT and 90.5\,mT have been measured with higher accuracy than reported previously \cite{inouye1998oof}. Guided by initial theoretical models we performed a restricted search at higher magnetic fields and located two additional Feshbach resonances, namely one $d$-wave resonance at 159.0\,mT and one $s$-wave resonance at 205.4\,mT.

The required hold times range from 25\,ms for the broadest $s$-wave resonances up to 4\,s for the very narrow $d$- and $g$-wave resonances. Except for the $s$-wave resonances, the typical width of the loss features is about 10\,$\mu$T to \,20\,$\mu$T, mostly determined by our magnetic field stability. We take the magnetic field value of maximal loss as the resonance position $B_0^{\rm exp}$. The magnetic field is calibrated close to each Feshbach resonance by means of RF spectroscopy on the $|1,1\rangle\rightarrow|1,0\rangle$ transition of Na. The asymmetry observed in the $s$-wave loss features in Fig.\,\ref{FRall}(k) and \ref{FRall}(\textit{l}) is caused by molecule association when ramping down the magnetic field for imaging.

\begin{table}
\caption{Overview of the experimentally and theoretically obtained results on Feshbach resonances in Na prepared in the lowest hyperfine substate $|F,m_F\rangle=|1,1\rangle$. Experimentally the position of maximum loss $B_0^{\rm exp}$ is given, which is determined by Gaussian fits to the loss features. The  errors reflect the one standard deviation statistical uncertainty in the magnetic field calibration, determined by the FWHM of the RF calibration spectra and from the profile fit. The middle columns shows the results for the Moerdijk and ABM models in terms of $B_c$ (see Sec.\,\ref{ABM}). The last two columns show the Feshbach resonance position $B_0$ and width $\Delta$ from coupled-channels (CC) calculation, based on refined Na$_2$ ground-state potentials (see Sec.\,\ref{CC}). The brackets in the last column give the exponent to the power ten. The quantum numbers $l, f, m_f$ characterize the bound state and are discussed in the text. Note that for the $l$=4 states the $m_f$ quantum numbers could not be assigned (see Sec.\,\ref{CC})} 
\label{FRlist}
\begin{ruledtabular}
\begin{tabular}{ccc| c | c c |cc }
  & & & Exp.  & Moerdijk & ABM & \multicolumn{2}{c}{CC} \\
$l$ & $f$ & $m_f$ & $B_0^{\rm exp}$(mT) &  $B_c$(mT) & $B_c$(mT) &  $B_0$(mT) & $\Delta$(mT) \\
\hline
 0 & 4 & 2 & 85.10(2) & 84.82  & 85.10  & 85.114 &  9.7[-4]\\
 0 & 2 & 2 & 90.51(4) & 90.81  & 90.52  & 90.517 &  0.104\\
 0 & 3 & 2 & 205.42(4)& 205.45 & 205.44 & 205.501 & 0.012  \\
\hline
 2 & 4 & 4 & 49.36(2) & 49.34  & 49.40 &  49.344 & 1.7[-4]\\
 2 & 4 & 3 & 53.66(2) & 53.57  & 53.63 &  53.650 & 3[-5]\\
 2 & 4 & 2 & 58.63(2) & 58.52  & 58.58 &  58.615 & 5[-6]\\
 2 & 4 & 1 & 64.48(3) & 64.41  & 64.47 &  64.477 & 3[-7]\\
 2 & 2 & 2 & 66.28(3) & 66.38  & 66.28 &  66.283 & 4[-7]\\
 2 & 4 & 0 & 71.56(1) & 71.54  & 71.60 &  71.556 & 2[-8] \\
 2 & 2 & 1 & 72.71(1) & 73.03  & 72.72 &  72.718 & 4[-7]\\
 2 & 3 & 3 & 159.00(3)& 159.02 & 159.13&  159.011 & 2[-4]\\
\hline
 4 & 3 &   & 50.80(2) & 50.76  & 50.80 & 50.775  &$<5[-7]$\\
 4 & 3 &   & 50.88(2) & 50.89  & 50.89 & 50.859 & $<5[-7]$\\
 4 & 3 &   & 51.09(2) & 51.12  & 51.09 & 51.065  &$<5[-7]$\\
\end{tabular}
\end{ruledtabular}
\end{table}

The measured position of the broadest $s$-wave resonance of 90.51(4)\,mT is in agreement with the previously reported value of 90.7(2.0)\,mT \cite{inouye1998oof}. We find the spacing between the lowest two $s$-wave resonances to be 5.41(5)\,mT, also in agreement with the earlier value of 5.4(1)\,mT \cite{inouye1998oof}. We note that preliminary data on part of the Na $d$-wave resonances have been reported in Ref.\,\cite{stan2005PhD}.

\section{Theory}\label{Theory}

\subsection{Hamiltonian}\label{intro hamiltonian}

Feshbach resonances occur when molecular states are resonant with two scattering atoms. For sodium atoms in the ground state $^2S_{1/2}$ the molecular states are related to the least-bound rovibrational levels of the singlet \Xstate~and triplet \astate~Born-Oppenheimer potentials. Their rovibrational levels are labeled by the vibrational and rotational quantum numbers $v$ and $l$ as well as the space-fixed projection $m_l$ of the orbital angular momentum along the magnetic field direction. The basis $|\sigma\rangle=|SM_SIM_I\rangle$ describes the remainder of the molecular wavefunction. The spins $\vec S$ and $\vec I$ are the total electron and nuclear spin of the molecule, respectively. In fact, for a pair of atoms $A$ and $B$ we have $\vec{S}=\vec{s}_A+\vec{s}_B$ and $\vec{I}=\vec{\imath}_A+\vec{\imath}_B$, where $\vec{s}_{A/B}$ and $\vec{\imath}_{A/B}$ are the electron and nuclear spin of each of the atoms. Projections are again defined along the magnetic field direction.

There is significant hyperfine interaction in Na and it is also useful to define the coupled basis $|(SI)f,m_f\rangle$, in which the total electronic and nuclear spins are coupled to $\vec{f}=\vec{S}+\vec{I}$, and $m_f$ is its projection. We will label the molecular states with quantum numbers $l$ and $m_f$, common to both bases, and if appropriate with $f$ or $M_S$.  For homonuclear dimers symmetrization ensures that
only states with $I+S+l$ even exist \cite{stoofsea1988}.

In order to accurately and quantitatively describe both bound states and scattering
processes, which include Feshbach resonances, we require the Hamiltonian for two sodium atoms in a magnetic field $B$ (see e.\,g.\,\cite{stoofsea1988,laue2002mfi}). Here, we use an extension that allows for a second-order spin-orbit interaction \cite{mies1996ebo} and a hyperfine-contact interaction that depends on the internuclear separation $R$. The effect of this second correction was recently studied for Rb$_2$ \cite{strauss2010}.  Finally, we have
\begin{eqnarray}
H &=& T + H_{\rm hf}(R) + H_Z
      + U_{\rm X}(R)P_{\rm X} + U_{\rm a}(R)P_{\rm a}
\nonumber\\
  &&\quad \quad + V_{SS}(\vec R)\,,
\label{ccham}
\end{eqnarray}
where $T=-\hbar^2\nabla^2/(2\mu)$ is the relative kinetic energy operator with reduced mass $\mu$,
\[
H_{\rm hf}(R)=\sum_{\alpha=A,B}
a_\alpha(R)\vec{s}_\alpha\cdot\vec{\imath}_\alpha/\hbar^2
\] is the
$R$-dependent hyperfine-contact interaction, and
\[ H_Z=\sum_{\alpha=A,B}
(g_{s\alpha}s_{z\alpha}+g_{i\alpha}i_{z\alpha}) \mu_B B/\hbar
\]
is the magnetic Zeeman interaction. In the limit $R\to\infty$, the $R$-dependent functions $a_\alpha(R)$ approach the atomic values $a_{\rm hf}$. The constants $g_{s\alpha}$, $g_{i\alpha}$, and $\mu_B$ are the sodium electron gyromagnetic ratio, sodium nuclear gyromagnetic ratio, and the Bohr magneton, respectively. All atomic constants are taken from Ref.\,\cite{arimondo1977edo} except for the atomic mass of sodium, which is obtained from Ref.\,\cite{aud03}.

The last two terms on the first line of Eq.~\ref{ccham} describe the Hamilton operator for the Born-Oppenheimer potentials $U_{\rm X}(R)$ and $U_{\rm a}(R)$.  The operators $P_{\rm X}$ and $P_{\rm a}$ project on spin basis functions $|SM_S IM_I\rangle$ with total electron spin $S=0$ and 1, respectively. Consequently, $P_{\rm X}+P_{\rm a}=1$. Finally, $V_{SS}(\vec R)$ is a weak but non-negligible effective electron-spin
electron-spin interaction that includes the second-order spin-orbit interaction and depends on the orientation of the internuclear axis $\vec R$ as well as the internuclear separation. Therefore, $V_{SS}(\vec R)$ is the only term in the Hamiltonian that mixes or couples partial waves $\vec l$. It will be defined more precisely in the Sec.\,\ref{CC}.

The Hamiltonian conserves the total angular momentum $\vec f+\vec l$ at zero magnetic field, and always conserves its projection $M=m_f+m_l$. Neglecting the potential $V_{SS}(\vec R)$, the Hamiltonian conserves $\vec f$ for $B=0$, while for finite field only the projection $m_f$ remains a good quantum number. For large magnetic field, where $H_Z$ is larger than $H_{\rm hf}(R)$, the total electron spin $S$ and its projection $M_S$ are approximately conserved. For sodium this corresponds to fields $B>30$\,mT.

\begin{table}
\caption{List of $l=0$ and 2 states for each vibrational singlet ($S=0$) and triplet ($S=1$) state, for which $M=m_f+m_l=2$ holds. The last two columns show the quantum numbers of the coupled basis, which holds for low magnetic fields.\label{listSandTstates}}
\begin{ruledtabular}
\begin{tabular}{c c || c c c c || c c }
$l$ & $m_l$ & $S$ & $M_S$ & $I$ & $M_I$ & $(SI)f$ & $m_f$ \\
\hline
0 & 0& 0 & 0   & 2 & 2 & (02)2 & 2 \\
\hline
0 & 0 & 1 & 1 & 3 & 1 & (13)4 & 2\\
  &   &   & 1 & 1 & 1 & (11)2 & 2\\
  &   &   & 0 & 3 & 2 & (13)3 & 2\\
  &   &   & -1& 3 & 3 & (13)2 & 2 \\
\hline
2 & 0& 0 & 0   & 2 & 2     &(02)2 & 2  \\
  & 1&   &     & 2 & 1     & (02)2 & 1   \\
  & 2&   &     & 2 & 0     &(02)2 & 0   \\
  & 2&   &     & 0 & 0     &(00)0 & 0   \\
\hline
2 & -2 & 1 & 1 & 3 & 3 & (13)4 & 4\\
  &-1  &   & 1 & 3 & 2 & (13)4 &3\\
  & 0  &   & 1 &3  & 1 & (13)4 &2\\
  & 1  &   & 1 &  3& 0 & (13)4 &1\\
  & 2  &   & 1 &  3& -1& (13)4 &0\\
  & 0  &   & 1 &  1& 1 & (11)2 &2\\
  & 1  &   & 1 &  1& 0 & (11)2 &1\\
  & 2 &   & 1   & 1 & -1    &(11)2 & 0  \\
  & -1&   & 0 & 3& 3& (13)3& 3\\
  & 0 &   & 0   & 3 & 2     &(13)3 & 2  \\
  & 1 &   & 0   & 3 & 1     & (13)3 & 1 \\
  & 2 &   & 0   & 3 & 0     & (13)3 & 0  \\
  & 1 &   & 0   & 1 & 1     & (11)1 & 1  \\
  & 2 &   & 0   & 1 & 0    & (11)1 & 0 \\
  & 0 &   & -1  & 3 & 3     & (13)2 & 2  \\
  & 1 &   & -1  & 3 & 2     & (13)2 & 1  \\
  & 2 &   & -1  & 3 & 1   & (13)2 & 0 \\
  & 2 &   & -1  & 1 & 1     & (11)0 & 0  \\
\end{tabular}
\end{ruledtabular}
\end{table}

We focus on ultra-cold collisions, with temperatures well below the $p$-wave centrifugal barrier. Hence, only $s$-wave collisions need to be considered. This also means that the $d$- and $g$-wave resonances can only be induced by the weak spin-spin interaction $V_{SS}(\vec R)$, which allows coupling between the $s$-wave continuum and $l=2$ and 4 molecular states, respectively. Moreover, for Na atoms in the $|1,1\rangle$ hyperfine state, $m_f=m_{f_A}+m_{f_B}=2$ and, therefore, we can focus on basis functions with $M=2$. The spin basis states $|\sigma\rangle$ for $l=0$ and 2 are listed in Table\,\ref{listSandTstates}. Also shown is an equivalent list in the coupled $|(SI)f,m_f\rangle$ basis. With the constraint $M=2$, there are five $l=0$ basis states, where only one has singlet character, whereas for $l=2$ there are four singlet and 18 triplet basis states.

\subsection{Asymptotic-bound-state model}\label{ABM}

In general, it is difficult to predict the location of Feshbach resonances, because of their sensitivity to details of the molecular potentials. This has lead to the development of models that require minimal knowledge of the molecular potentials (see e.\,g.\,\cite{vogels1998dmf,houbiers1998eai,verhaar2009psp,hanna2009pof}), at the expense of accuracy and applicability. A powerful, yet computationally light description of the near threshold molecular spectrum is the asymptotic bound-state model (ABM)
\cite{wille2008eau,tiecke2010abs}. It builds on an earlier model by Moerdijk \emph{et al.} \cite{moerdijk1995riu} for homonuclear systems, which was extended by Stan \emph{et al.} \cite{stan2004oof} for heteronuclear systems. By using binding energies and wavefunction overlaps between singlet and triplet rovibrational levels as fit parameters it circumvents the need of any knowledge of the molecular potentials while allowing for a high degree of accuracy once the position of a few Feshbach resonances is known. This type of modeling is closely related to the well developed deperturbation theories of molecular spectra, see e.\,g.\ K$_2$ \cite{lisdat2001}.

In the ABM model we neglect the $R$ dependence of the hyperfine constants and set $V_{SS}(\vec R)$ to zero in the Hamiltonian of Eq.~\ref{ccham}. We then define the molecular wavefunctions $|vl,\sigma\rangle=|\psi_v^{S,l}\rangle|\sigma\rangle$, where
$|\psi_v^{S,l}\rangle$ describe vibrational and rotational wavefunctions of the $S=0$ singlet or $S=1$ triplet potential, and formally
construct the Hamiltonian matrix $\cal H$ with matrix elements
\[
      \langle v'l',\sigma'| H | vl,\sigma\rangle
\]
for a limited set of near-threshold bound states of the two Born-Oppenheimer potentials and spin states $|\sigma\rangle$ consistent with total projection quantum number $M$.

For our homonuclear system we find
\begin{eqnarray}\label{ABMmatrix}
  {\cal H}_{v'l'\sigma',vl\sigma} &=&
   \varepsilon^{S,l}_v\delta_{vv'}\delta_{ll'}\delta_{\sigma\sigma'} \\
    &&  + \mu_B B (g_s M_S + g_i M_I) \delta_{vv'}\delta_{ll'}\delta_{\sigma\sigma'}
             \nonumber\\
    &&  + a_{\rm hf} \delta_{ll'}\,\eta^{SS'}_{vv'}(l)\,
       \langle \sigma' | \sum_{\alpha} \vec s_\alpha\cdot \vec\imath_\alpha |\sigma\rangle/\hbar^2
         \nonumber
\end{eqnarray}
where $\delta_{ij}$ is the Kronecker delta, $\varepsilon^{S,l}_{v}$ is the energy of the rovibrational level ($v$, $l$) of the singlet or triplet potential, and $\eta^{SS'}_{vv'}(l) =\langle \psi_v^{S,l}|\psi_{v'}^{S',l}\rangle$ is the overlap integral of two rovibrational wavefunctions. Clearly, for $S=S'$ we have $\eta^{SS}_{vv'}(l)=\delta_{vv'}$. The operator $\sum_{\alpha} \vec s_\alpha\cdot \vec\imath_\alpha$ does not commute with the total electron spin $S$. In fact, it is the only part of $\cal H$ that couples singlet and triplet levels.

The eigenvalues of the Hamiltonian $\cal H$ are the molecular bound states. We can label these levels by $M$, $m_f$, and $l$ as well as their $B=0$ quantum number $f$. As function of magnetic field some of these molecular bound states cross the field-dependent $|1,1\rangle+|1,1\rangle$ dissociation limit. Such a crossing position is denoted by $B_c$, which principally differs from the Feshbach resonance position $B_0$ (as defined in Eq.\,\ref{scatteringlength}) because finite coupling between the molecular bound states and the scattering channel shifts the resonance position \cite{chin2010fri}. For the goal of the ABM of giving a simplified picture the shifts are not important in the case of Na$_2$. We adjust the energies $\varepsilon^{S,l }_{v}$ and overlaps $\eta^{SS'}_{v,v'}(l)$ to fit the molecular spectrum to the observed Feshbach resonances.

\subsubsection{Moerdijk model}

We first neglect singlet-triplet coupling by applying the following approximation in Eq.~\ref{ABMmatrix}:
\[
\sum_\alpha \vec s_\alpha \cdot \vec i_\alpha \approx \frac{1}{2} \vec S \cdot \vec I,
\]
such that the hyperfine interaction commutes with the total electron spin $S$. We refer to this approximation of ABM as the ``Moerdijk model'' \cite{moerdijk1995riu}. In the Moerdijk model the only fit parameters are the energies $\varepsilon^{S,l }_{v}$, as the overlap parameters $\eta^{SS'}_{vv'}(l)$ do not appear.

\begin{figure}
\includegraphics[width=8.5cm]{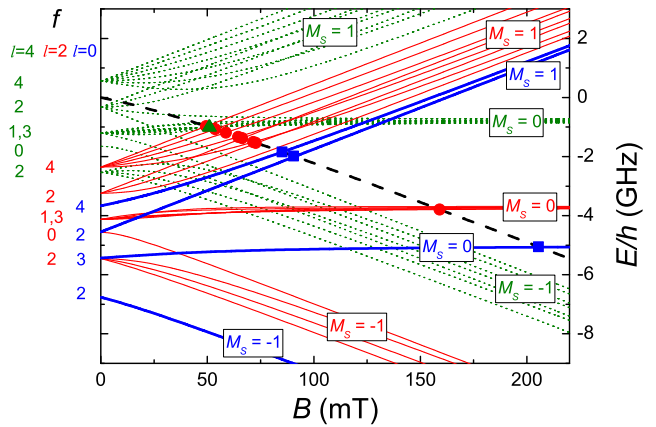}
\caption{(color online) Moerdijk model of the $v_a=14$ triplet vibrational level with $M=2$, and $l=0$ (thick blue lines), $l=2$ (thin red lines) and $l=4$ (dotted green lines) rotational states. The dashed black line is the atomic $|1,1\rangle+|1,1\rangle$ threshold. The observed Feshbach resonances are indicated by the filled squares ($s$-wave resonances), circles ($d$-wave resonances) and triangles ($g$-wave resonances). Note that the three $g$-wave resonances overlap with one $d$-wave resonance in this graph. On the left side of graph the $B=0$ quantum number $f$ is indicated for the different $l$ states, whereas inside the graph the approximate quantum number $M_S$ is denoted.}\label{Moerdijk_triplet}
\end{figure}

We find that all the observed Feshbach resonances can be described by a single vibrational triplet level below the $|1,1\rangle+|1,1\rangle$ dissociation threshold. In fact it is the $v_a=14$ level with $l=0$, 2 and 4 \cite{laue2002mfi}. From a least-squares fit including all resonances we obtain for the rotational structure of this vibrational level three fitted energies: $\varepsilon^{1,0}_{14}/h=-4976$\,MHz, $\varepsilon^{1,2}_{14}/h=-3679$\,MHz and $\varepsilon^{1,4}_{14}/h=-765$\,MHz, where the energies are given with respect to the $|1,1\rangle+|1,1\rangle$ dissociation threshold.
The resulting molecular spectrum is shown in Fig.\,\ref{Moerdijk_triplet} and demonstrates that the Moerdijk model describes the spectrum of Feshbach resonances already quite well. The corresponding Feshbach resonance positions are given in Table\,\ref{FRlist} under ``Moerdijk''. The deviations with the experimental values are at most 0.3\,mT, much smaller than the spacing between the resonances, and allows for an unambiguous labeling of the molecular states causing the resonances. The observation of three $g$-wave Feshbach resonances near 51\,mT close to the $d$-resonances can only be explained by the presence of $f=3$ ($M_S=0$) states. However, an $m_f$-assignment is ambiguous as singlet-triplet coupling and spin-spin interaction plays a significant role for these nearly degenerate states. Thus no $m_f$ label is given in Table\,\ref{FRlist} for these cases (see Sec.\,\ref{CC} for further discussion).

In Fig.\,\ref{Moerdijk_triplet} the level structure is continued beyond the $|1,1\rangle+|1,1\rangle$ dissociation threshold into the continuum. All these levels have finite life times. The group of $l=4$ levels, which for all $B$ have an energy above the dissociation limit, could give rise to shape resonances.

\subsubsection{ABM}

To improve the description of the Feshbach data one has to include the singlet-triplet coupling and consider all rovibrational levels of the singlet and triplet potentials with binding energies less than a few times the Zeeman and hyperfine splitting. For Na$_2$ these are the $v_a=14$, 15 and $v_X=64$, 65 rovibrational levels \cite{samuelis2000cac,laue2002mfi,araujo2003tcp}.

There are eight fit parameters (four energies and four overlap integrals) for $l=0$, for which there are only three resonances to constrain their values. If we assume that the energies and overlaps of different $l$ states are independent, we have in total 32 fit parameters to fit only 14 resonances. Consequently these fits are underconstrained. In principle, however the number of fit parameters can be reduced by noting that binding energies and overlaps for $l>0$ follow from that for $l=0$ by using model potentials or the accumulated phase method \cite{wille2008eau,tiecke2010abs}. However, as we want to use the ABM model in its simplest form we fit the $\varepsilon^{S,l}_{v}$ as well as the $\eta^{SS'}_{vv'}(l)$ parameters independently.

We can improve upon the fit based on the Moerdijk model. We have identified a class of states that are mostly insensitive to singlet-triplet mixing. The locations of the resonances caused by $f=4$ states only depend to well within the experimental uncertainty on the $\varepsilon^{1,l}_{14}$ parameters. Thus, we can improve the level energies compared to those obtained within the Moerdijk model, as we can exclude resonances that are sensitive to singlet-triplet mixing. We find $\varepsilon^{1,0}_{14}/h=-4991(1)$\,MHz and $\varepsilon^{1,2}_{14}/h=-3682(2)$\,MHz.
The Feshbach resonance positions caused by these $f=4$ states are given in Table\,\ref{FRlist} under ``ABM''. The calculated positions of the $f=4$ $d$-wave resonances deviate only by 0.05\,mT or less from the experimental values.

To apply the ABM model to Feshbach resonances that are sensitive to the $v_X=64$ and 65 singlet states, and indirectly to the $v_a=15$ triplet state, we include previous experimental data to reduce the number of fit parameters. First, the $\varepsilon^{1,0}_{15}$ parameter can be extracted from photoassociation spectroscopy measurements \cite{araujo2003tcp} using a state that is insensitive to singlet-triplet mixing. One obtains $\varepsilon^{1,0}_{15}/h=+2014(10)$\,MHz. From magnetic field dependent molecular beam spectroscopy experiments on an avoided crossing between the $v_a=14$ ($f=2$, $M_S=-1$) and $v_X=64$ ($f=2$, $M_S=0$) states around $B=170$\,mT \cite{laue2002mfi} we extract $\varepsilon^{0,0}_{64}/h\approx-11$\,GHz and $\eta^{01}_{64,14}(0)\approx0.85$.

We then find a satisfactory fit for all $s$-wave resonances with the values 0.47, 0.79 and 1.0 for the remaining three overlap parameters $\eta^{01}_{64,15}(0)$,  $\eta^{01}_{65,14}(0)$ and $\eta^{01}_{65,15}(0)$. The fitted $s$-wave resonance positions are given in Table\,\ref{FRlist}. The fit is not unique, highlighting the difficulty of using a deperturbative ABM approach for this system.

For the $d$-wave resonances we follow a similar approach, extracting $\eta^{01}_{64,14}(2)\approx0.85$ from Ref.\,\cite{laue2002mfi}, and using the rotational splitting from Ref.\,\cite{samuelis2000cac}: $\varepsilon^{0,2}_{64}-\varepsilon^{0,0}_{64}=h\times1556(38)$\,MHz,
$\varepsilon^{0,2}_{65}-\varepsilon^{0,0}_{65}=h\times504(38)$\,MHz and
$\varepsilon^{1,2}_{15}-\varepsilon^{1,0}_{15}=h\times465(34)$\,MHz. With the three remaining $d$-wave resonances we obtain $\eta^{01}_{64,15}(2)\approx0.44$ and $\eta^{01}_{65,14}(2)\approx0.47$ and calculate the $d$-wave resonances position (see Table\,\ref{FRlist}).

Finally, for the three $g$-waves resonances, which are caused by a group of $l=4$, $f=3$ ($M_S=0$) states, we only find a weak dependence on the $v_X=64$, 65 and $v_a=15$ levels. We use the rotational splitting
$\varepsilon^{0,4}_{64}-\varepsilon^{0,0}_{64}=5064(45)$\,MHz from Ref.\,\cite{samuelis2000cac}, and find the $l=4$ level of the $v_X=65$ and $v_a=15$ levels assuming the ratio between the $l=2-l=0$ and $l=4-l=0$ rotational splitting of the $v_X=64$ level. From a least-square fit we obtain the energy parameter of the $v_a=14$ triplet state, $\varepsilon^{1,4}_{14}/h=-777(2)$\,MHz, and calculated the $g$-wave resonance positions (see Table\,\ref{FRlist}).

In summary, we can extract within the ABM the binding energies of the $v_a=14$ level by selecting those resonances that are not sensitive to singlet-triplet coupling. To obtain the other resonances we need to include spectroscopy data to restrict the number of fit parameters. However, the obtained set of fit parameters is not unique. For instance, we have found that $\eta^{01}_{65,15}(l)$ and $\varepsilon^{0,l}_{65}$ are correlated, thus choosing a different $\eta^{01}_{65,15}(l)$ would immediately result in a different $\varepsilon^{0,l}_{65}$. In fact, from the $l=0$ $v_a=15$ and $v_X=65$ wavefunctions obtained by the coupled-channel calculation (see Sec.\ref{CC}) we find that $\eta^{01}_{65,15}(0)$=0.60, which significantly deviates from our fit.

\subsection{Coupled-channels calculation\label{CC}}

The description based on the Moerdijk and ABM models agrees with the experimental resonance locations to well within one percent. However, the ABM is only valid over the limited range of energies spanned by the included vibrational levels and can not be used directly without further assumptions to find scattering lengths. Any extrapolation out of this limited energy region is not reliable. Additionally, the approach needs assumptions on other parameters like overlap integrals to obtain a number of free parameters that is at least equal to the number of observations. Thus the overall validity of this approach should be checked by a more general theoretical model.

We use a coupled-channels approach based on the Hamiltonian in Eq.\,\ref{ccham} with $R$-dependent interaction potentials and coupling terms. The $R$-dependent functions are found in a fit to all known Feshbach resonances and existing spectroscopic data of more deeply bound rovibrational levels of the \Xstate~and \astate~ potentials \cite{elbs1999oot,samuelis2000cac,laue2002mfi,fatemi2002ugs,araujo2003tcp}. Data from conventional spectroscopy by Kush and Hessel \cite{kush1978} and by Barrow \textit{et al.} \cite{barrow1984} for the ground state \Xstate~potential, and by Li Li \textit{et al.} \cite{li1985} for the lowest triplet state \astate~potential are also included. The fits based on the coupled-channels solutions will then lead to a consistent description of the sodium dimer that is valid over the full depth of the \Xstate~and \astate~potentials and will be a well justified system to predict scattering properties or dynamics like spin exchange.

The hyperfine-contact interaction and $V_{SS}(\vec R)$ in Eq.\,\ref{ccham} require further discussion. The hyperfine interaction $a_{\alpha}(R)$ is $R$ dependent and accounts for the electronic distortions of one atom by the other. We use the simple Ansatz
\begin{equation}
\label{eq:hfs}
a_{\alpha}(R)=a_{\alpha,\rm hf}
         \left(1+ \frac{c_f}{e^{(R-R_0)/\Delta R} +1}\right)\,,
\end{equation}
which represents a switching function around $R_0$ with width $\Delta R$ to go from the atomic value $a_{\alpha,\rm hf}$ at large $R$ to the molecular value $a_{\alpha,\rm hf}(1+c_f)$ at small $R$. For our homonuclear diatom we have $a_{A}(R)=a_{B}(R)$. The parameter values are determined by the fitting process.

The effective spin-spin interaction, which couples the partial waves $l$, is given by
\begin{equation}
V_{SS}(\vec R)=\frac{2}{3}\lambda(R)(3S_Z^2-S^2)\,,
\end{equation}
where the operator $S_Z$ is the total electron spin projection operator along the internuclear axis $Z$ and
\begin{equation}
\label{lambda} \lambda(R) = -\frac{3}{4} \alpha^2
\left(\frac{1}{R^3}+a_\mathrm{SO} e^{-b_{\rm SO}R}\right)\,.
\end{equation}
The function $\lambda(R)$ consists of two terms. The first represents the magnetic dipole-dipole interaction, the second is modeling the second order spin-orbit contribution. When $\lambda(R)$ and $R$ are given in atomic units, $\alpha$ is the fine structure constant. For sodium this Ansatz was first used in Ref.~\cite{araujo2003tcp}, where they found that a non-negligible $a_\mathrm{SO}$ is needed to describe the photoassociation data.

We represent the \Xstate~and \astate~Born-Oppenheimer potentials using a separate functional form for three regions of the internuclear separation: a short range repulsive wall for $R<R_\mathrm{inn}$, an asymptotic long range region for $R>R_\mathrm{out}$, and an intermediate deeply bound region in between. The analytic form of the potentials in
the intermediate region, $U_{\rm IR}(R)$, is the finite power expansion
\begin{equation}
\label{eq:uanal} U_{\rm IR}(R)=\sum_{i=0}^{n}a_i\,\xi^i(R)\,,
\end{equation}
in the nonlinear $R$-dependent function $\xi(R)$ given by
\begin{equation}
\label{eq:xv} \xi(R)=\frac{R - R_m}{R + b\,R_m}\,.
\end{equation}
Here, the upper limit $n$ and the $a_i$ are fit parameters and we choose $b$ and $R_m$ such that only a small number of $a_i$ are needed. The separation $R_m$ is chosen near the value of the equilibrium separation. For $R < R_\mathrm{inn}$ the potential is extrapolated with
\begin{equation}
\label{eq:rep}
  U_\mathrm{SR}(R)= u_1 + u_2/R^{s}\,.
\end{equation}
where $u_1$ and $u_2$ are adjusted to create a continuous and differentiable transition at $R_{\rm inn}$. We use $s=6$ for both \Xstate~and \astate~potentials.

For $R > R_{\rm out}$ we adopt the long range form
\begin{equation}
\label{eq:lrexp} U_{\mathrm {LR}}(R)=-C_6/R^6-C_8/R^8
-C_{10}/R^{10}\mp E_{\mathrm{exch}}(R)\,,
\end{equation}
where the exchange contribution changes the sign between singlet and triplet state and the form is given by \cite{exc}
\begin{equation}
\label{eq:exch} E_{\mathrm{exch}}(R)=A_{\mathrm{ex}} R^\gamma
\exp(-\beta R)\,.
\end{equation}

\begin{table}[h]
\fontsize{8pt}{8pt}\selectfont
\caption{Parameters of the analytic representation of the \Xstate\ state potential. The energy reference is the dissociation asymptote. Parameters with an asterisk $^\ast$ ensure smooth continuous extrapolation of the potential at $R_{\rm inn}$.}
\label{tab:Xpot}
\vspace{2mm}
\begin{tabular*}{1.0\linewidth}{@{\extracolsep{\fill}}lr}
\hline
\multicolumn{2}{c}{$R < R_\mathrm{inn}= 2.181$ \AA} \\
\hline
$u_1$* & $-0.785318644\times 10^{4}$ \wn \\
$u_2$* & $ 0.842586535\times 10^{6}$ \wn \AA $^{6}$ \\
\hline
\multicolumn{2}{c}{$R_\mathrm{inn} \leq R \leq R_\mathrm{out}=  11.00$ \AA} \\
\hline
$b$ & $-0.140$ \\
$R_m$ & $3.0788576$ \AA \\
 $a_{0}$ & $-6022.04193$ \wn\\
 $a_{1}$ & $-0.200727603516760356\times 10^{ 1}$ \wn\\
 $a_{2}$ & $ 0.302710123527149044\times 10^{ 5}$ \wn\\
 $a_{3}$ & $ 0.952678499004718833\times 10^{ 4}$ \wn\\
 $a_{4}$ & $-0.263132712461278206\times 10^{ 5}$ \wn\\
 $a_{5}$ & $-0.414199125447689439\times 10^{ 5}$ \wn\\
 $a_{6}$ & $ 0.100454724828577862\times 10^{ 6}$ \wn\\
 $a_{7}$ & $ 0.950433282843468915\times 10^{ 5}$ \wn\\
 $a_{8}$ & $-0.502202855817934591\times 10^{ 7}$ \wn\\
 $a_{9}$ & $-0.112315449566019326\times 10^{ 7}$ \wn\\
$a_{10}$ & $ 0.105565865633448541\times 10^{ 9}$ \wn\\
$a_{11}$ & $-0.626929930064849034\times 10^{ 8}$ \wn\\
$a_{12}$ & $-0.134149332172454119\times 10^{10}$ \wn\\
$a_{13}$ & $ 0.182316049840707183\times 10^{10}$ \wn\\
$a_{14}$ & $ 0.101425117010240822\times 10^{11}$ \wn\\
$a_{15}$ & $-0.220493424364290123\times 10^{11}$ \wn\\
$a_{16}$ & $-0.406817871927934494\times 10^{11}$ \wn\\
$a_{17}$ & $ 0.144231985783280396\times 10^{12}$ \wn\\
$a_{18}$ & $ 0.379491474653734665\times 10^{11}$ \wn\\
$a_{19}$ & $-0.514523137448139771\times 10^{12}$ \wn\\
$a_{20}$ & $ 0.342211848747264038\times 10^{12}$ \wn\\
$a_{21}$ & $ 0.839583017514805054\times 10^{12}$ \wn\\
$a_{22}$ & $-0.131052566070353687\times 10^{13}$ \wn\\
$a_{23}$ & $-0.385189954553600769\times 10^{11}$ \wn\\
$a_{24}$ & $ 0.135760501276292969\times 10^{13}$ \wn\\
$a_{25}$ & $-0.108790546442390417\times 10^{13}$ \wn\\
$a_{26}$ & $ 0.282033835345282288\times 10^{12}$ \wn\\
\hline
\multicolumn{2}{c}{$R_\mathrm{out} < R$}\\
\hline
${U_\infty}$ & 0.0 \wn \\
${C_6}$ & 0.75186131$\times 10^{7}$ \wn\AA$^6$ \\
${C_{8}}$ & 0.1686430$\times 10^{9}$ \wn\AA$^8$ \\
${C_{10}}$ & 0.3081961$\times 10^{10}$ \wn\AA$^{10}$ \\
${A_{\rm ex}}$ & 0.40485835$\times 10^{5}$ \wn\AA$^{-\gamma}$ \\
${\gamma}$ & 4.59105 \\
${\beta}$ & 2.36594 \AA$^{-1}$ \\
\hline
\hline
\multicolumn{2}{c}{Derived constants:} \\
\hline
\multicolumn{2}{l}{equilibrium distance:\hspace{1.6cm} $R_e^X$= 3.07895(5) \AA} \\
\multicolumn{2}{l}{electronic term energy:\hspace{0.3cm} $T_e^X$=$-D_e^X$= -6022.0420(40) \wn}\\
\hline
\end{tabular*}
\end{table}

\begin{table}[h]
\fontsize{8pt}{8pt}\selectfont
\caption{Parameters of the analytic representation of the potential curve of the \astate~state. The energy reference is the dissociation asymptote. Parameters with an asterisk $^\ast$ ensure smooth continuous extrapolation of the potential at $R_{\rm inn}$.}
\label{tab:apot}\vspace{2mm}
\begin{tabular*}{1.0\linewidth}{@{\extracolsep{\fill}}lr}
\hline
\multicolumn{2}{c}{$R < R_\mathrm{inn}= 4.2780$ \AA} \\
\hline
$u_1$* & $-0.2435819\times 10^{3}$ \wn \\
$u_2$* & $ 0.1488425\times 10^{7}$ \wn \AA $^{6}$ \\
\hline
\multicolumn{2}{c}{$R_\mathrm{inn} \leq R \leq R_\mathrm{out}=  11.00$ \AA} \\
\hline
$b$ & $-0.40$ \\
$R_m$ & $5.149085$ \AA \\
 $a_{0}$ & $-172.90517$ \wn\\
 $a_{1}$ & $ 0.355691862122135882\times 10^{ 1}$ \wn\\
 $a_{2}$ & $ 0.910756126766199941\times 10^{ 3}$ \wn\\
 $a_{3}$ & $-0.460619207631179620\times 10^{ 3}$ \wn\\
 $a_{4}$ & $ 0.910227086296958532\times 10^{ 3}$ \wn\\
 $a_{5}$ & $-0.296064051187991117\times 10^{ 4}$ \wn\\
 $a_{6}$ & $-0.496106499110302684\times 10^{ 4}$ \wn\\
 $a_{7}$ & $ 0.147539144920038962\times 10^{ 5}$ \wn\\
 $a_{8}$ & $-0.819923776793683828\times 10^{ 4}$ \wn\\
\hline
\multicolumn{2}{c}{$R_\mathrm{out} < R$}\\
\hline
${U_\infty}$ & 0.0 \wn \\
${C_6}$ & 0.75186131$\times 10^{7}$ \wn\AA$^6$ \\
${C_{8}}$ & 0.1686430$\times 10^{9}$ \wn\AA$^8$ \\
${C_{10}}$ & 0.3081961$\times 10^{10}$ \wn\AA$^{10}$ \\
${A_{\rm ex}}$ & 0.40485835$\times 10^{5}$ \wn\AA$^{-\gamma}$ \\
${\gamma}$ & 4.59105 \\
${\beta}$ & 2.36594 \AA$^{-1}$ \\
\hline
\hline
\multicolumn{2}{c}{Derived constants:} \\
\hline
\multicolumn{2}{l}{equilibrium distance:\hspace{1.6cm} $R_e^a$= 5.1431(10) \AA} \\
\multicolumn{2}{l}{electronic term energy:\hspace{0.3cm} $T_e^a$=$-D_e^a$= -172.909(40) \wn}\\
\hline
\end{tabular*}
\end{table}

We start the optimization process of the potentials with functions known from earlier work \cite{laue2002mfi} in iteration loops containing two steps. First, the location of the Feshbach resonances and the photoassociation data from Ref.\,\cite{fatemi2002ugs} are used to fit the long range behavior of Eq.\,\ref{eq:lrexp}. With the new potential functions from this first step all pure singlet and triplet rovibrational levels related to the Feshbach and photoassociation data are calculated and the results are added to the spectroscopic data of Refs.\,\cite{kush1978,barrow1984,li1985,laue2002mfi}. Then the second fitting step yields the improved potentials including all regions in $R$. With this result the first fit was repeated from which again better rovibrational levels will be derived for the overall potential fit. This iteration was continued until a self consistent status was reached. Because the $g$-wave resonances require long computation times, they are not included in the fit, but in each iteration loop the consistency of these data was checked.

In the fit the position of the Feshbach resonances are determined by the maximum of the elastic collision rate at a collision energy between 1 and 2\,$\mu$K, corresponding to the average temperature in the ensemble prepared during the experiment. This is not necessarily the correct position, because the observed loss of atoms is determined by three-body losses. We believe, however, that the difference between the resonance position defined by the experiment and that of the elastic resonance is not more than the experimental accuracy of $\lesssim0.01$\,mT.

The final potential parameters are given in Table\,\ref{tab:Xpot} and \ref{tab:apot} for the \Xstate~and \astate~states, respectively. The large number of digits is only given in order to create potentials with the numerical accuracy as applied in the fitting process. It does not reflect the physical accuracy of the potential. Most of the resonance positions are described well within the experimental uncertainty, see Table~\ref{FRlist}, and thus this approach yields a significant improvement compared to the ABM model. Additionally, the photoassociation data are reproduced with a standard deviation of $\sigma=0.82$ using the measurement uncertainties for weights. The fit shows that the $R$-dependence of the hyperfine coupling is given by $c_f=-0.029$, $R_0=11.0$ a.u.\,and $\Delta R=1.0$ a.u.\,, where the amplitude $c_f$ is dimensionless and we estimate its uncertainty as 30\%. $R_0$ is chosen such that the variation of the hyperfine coupling is mainly around the minimum of the triplet state, but certainly this choice is a bit arbitrary and different choices will alter the amplitude $c_f$. Nevertheless, we find that the hyperfine structure decreases from the atomic value by a few percent for $R<R_0$. Similarly, we varied the parameters for the effective spin-spin interaction and obtained $a_{SO}= -1.56$ a.u.\,and $b=1.0$ a.u.\,, where $b$ is chosen as in Ref.\,\cite{fatemi2002ugs} and the value of $a_{SO}$ has an estimated uncertainty of 100\%. It is almost a factor two smaller than in the above reference, but we should note the trend that the second order spin-orbit interaction slightly reduces the spin-spin interaction of the atomic pair in the molecule.

Jones \textit{et al.} \cite{jones1996} reported direct determination of the dissociation energy $D_0$ of Na$_2$ and found $D_0/(hc)$=5942.6880(39)\,\wn. We included this value in the fitting process and the model potential of \Xstate~reproduces this energy by 5942.6913\,\wn, thus in agreement with observation. The dissociation energy of the triplet state \astate~could not be improved, because it is only determined by the less precise spectroscopic data of Ref.\,\cite{li1985}. The derived values for $D_e$, which is the well of the Born-Oppenheimer potential and differs from $D_0$ by the zero-point energy, are given in Table\,\ref{tab:Xpot} and \ref{tab:apot}.

\begin{figure}
\includegraphics[width=8.5cm]{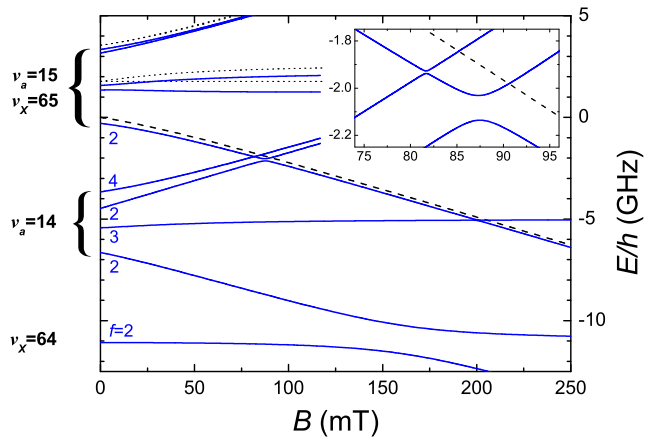}
\caption{(color online) Magnetic field dependence of the least bound $l=0$ singlet and triplet states with $M=2$. The dashed black line represents the dissociation threshold of two atoms in the $|1,1\rangle$ state, whereas the dotted black lines represent thresholds for asymptotes $F=1+F=2$ and $F=2+F=2$ with $M=2$. The inset zooms into the magnetic field region around the $s$-wave resonances at 85.1\,mT and 90.5\,mT. \label{overview}}
\end{figure}

In Fig.\,\ref{overview} the resulting molecular spectrum near the dissociation threshold is shown for $l=0$ states. This spectrum can be compared with that of Fig.\,\ref{Moerdijk_triplet} to observe the influence of the singlet levels via the singlet-triplet coupling. The most striking feature is the broad avoided crossing between the $v_a=14$ ($M_S=-1$) and $v_X=64$ ($M_S=0$) levels around 170\,mT, which was already investigated by Laue \textit{et al.} \cite{laue2002mfi}, and used in the ABM fitting procedure (see Sec.\,\ref{ABM}). The effect of the $v_X=65$ level is more subtle. The near-degenerate $v_X=65$ and $v_a=15$ levels are strongly mixed, leading to a significant singlet character of the $v_a=15$ triplet level. As a consequence, the $v_a=15$ $M_S=-1$ state can couple to the $v_a=14$ $M_S=1$ state, leading to avoided crossings at the intersects of these triplet states. This is highlighted in the inset of Fig.\,\ref{overview}, showing the region around the 85.1\,mT and 90.5\,mT $s$-wave resonances. The first avoided crossing is narrow and the effect on the position of the resonance caused by the $f=4$ state is negligible. However, the second avoided crossing is much broader and shifts the position of the resonance caused by the $f=2$ state by about 0.6\,mT.

The widths of Feshbach resonances are determined by the coupling of the bound levels to the continuum and thus directly calculable within the coupled-channels approach. Table\,\ref{FRlist} also lists the widths $\Delta$ defined according to Eq.\,\ref{scatteringlength} for elastic collisions with incoming $s$-wave. Only two of them, at 90.5\,mT and 205.4\,mT, are wide enough for use in an experiment with magnetic tunability. The others are so narrow that the hold time had to be large for detecting any enhanced atom loss. In fact, only because the applied magnetic field is fluctuating across the resonance, do these resonances become observable. In this situation the actual resonance width is reflected in the experimentally required hold time to observe the loss feature, and indeed we notice a clear correlation. We did search for the $f=2$ $d$-wave resonance that crosses the atomic threshold around 80\,mT (see Fig.\,\ref{Moerdijk_triplet}), but even a hold time of 10\,s did not result in a loss feature. According to the coupled-channels calculation the width of this particular resonance is even a factor of five smaller than the narrowest observed $d$-wave resonance at 71.6\,mT.

\begin{figure}
\includegraphics[width=7cm]{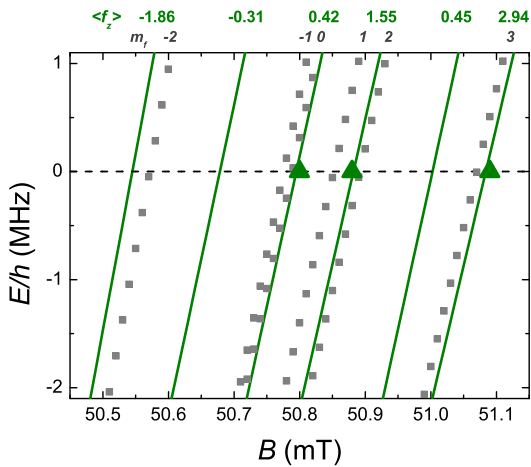}
\caption{(color online) Coupled-channels results for the group of $l=4$ bound states leading to the three observed $g$-wave resonances (green triangles), with (green solid lines) and without (grey squares) coupling between the different $m_f$ states. For the coupled states the expectation value of the operator $\hat{f}_z$ is given on top of the graph, explicitly showing that $m_f$ is not a good quantum number for these states. Energy is defined relative to the $|1,1\rangle$+$|1,1\rangle$ dissociation threshold.\label{gwaves}}
\end{figure}

Finally, we focus on the group of $l=4$ $f=3$ ($M_S=0$) states that causes the observed $g$-wave resonances. As stated earlier, for these states $m_f$ (or $m_l$) is not a good quantum number because of strong mixing between the different states. The molecular spectrum near the dissociation threshold is shown in Fig.\,\ref{gwaves}, in which also the expectation value $\langle\hat{f}_z\rangle$ is given. For comparison, the results of a calculation that excludes coupling between $m_f$ states are shown. Both the order of and spacing between the states are affected by the coupling.

\section{Conclusion}\label{concl}

We have observed 14 Feshbach resonances for ultracold Na in the absolute hyperfine ground state. The improvement on the field positions of the resonances, now better than 0.04\,mT, is very significant. Additionally, the newly found $g$-wave resonances extends the knowledge of the asymptotic bound structure of Na$_2$. A coupled-channels calculation describes most of the resonance position to well within the experimental uncertainty. The $R$-dependent hyperfine interaction has a noticeable effect on the magnetic field location of the Feshbach resonance. If we use the atomic hyperfine value for all internuclear separations we find that the resonances shifts up to 0.05\,mT, which is larger than our experimental uncertainty. With the new model potentials we derive for the location of the $s$-wave resonance in the $|1,-1\rangle+|1,-1\rangle$ entrance channel  120.75(0.5)\,mT, in agreement with the experimental value of 119.5(2.0)\,mT \cite{stenger1999sei}.

Our Feshbach spectroscopy data result in a precision of about 0.3\,MHz in the frequency scale for the weakly bound states closest to the atomic asymptote. Thus they complement the photoassociation data from the experiment by Fatemi \textit{et al.} \cite{fatemi2002ugs}, which have a precision of about 15\,MHz. Both data sets are strongly related to the triplet state, because the singlet-triplet mixing in Na$_2$ is weak as can be seen from the accuracy with which the Moerdijk model can describe the position of the Feshbach resonances. The combined evaluation of all existing data gives an accurate description of the \astate~potential down to about 820\,GHz where the measurements of the bound states by Ref.\,\cite{fatemi2002ugs} stop. Further down to its minimum only less precise and sparse data exist from the spectroscopy by Li Li \textit{et al.} \cite{li1985}.

Table\,\ref{levels} gives an overview of the most weakly bound level energies for the Moerdijk, ABM and coupled-channels models. In general there is fair agreement between ABM and the coupled-channels approach. One finds significant deviations for levels that are only weakly linked to the observed Feshbach resonances.

\begin{table}
\caption{Level energies $\varepsilon^{S,l}_{v}/h$ in MHz obtained from the different models for the two last bound states of the \Xstate~and \astate~potentials. The energy reference is the atomic pair asymptote $|1,1\rangle + |1,1\rangle$. The label * in the column ABM indicates that the value was estimated from other spectroscopic sources and not only from Feshbach spectroscopy, while the label $\dag$ denotes that more precise energy differences of rotational levels were used (see Sec.\,\ref{ABM}).} 
\label{levels}
\begin{ruledtabular}
\begin{tabular}{ccc| c |  c |c }
$X/a$ &  $v$&$l$&   Moerdijk & ABM & {CC} \\
\hline
 $a$ & 14 & 0 & -4976 & -4991(1)  & -4991.5(1.0)  \\
  $a$ &14 & 2 & -3679 & -3682(2)  & -3681.4(1.0)  \\
$a$ & 14 & 4 &  -764 & -777(2) & -779.8(1.0)  \\
\hline
 $a$ &15 & 0 &    & +2014(10)*  & +2033(15)\\
  $a$ &15 & 2 &   & +2479*$\dag$  & +2358(20)\\
\hline
  $X$ &64 & 0 &   & -11000*  & -10844(30)\\
  $X$ & 64 & 2 &  & -9444*$\dag$   & -9283(30)\\
  $X$ &64 & 4 &   & -5936*$\dag$   & -5749(40)\\
\hline
  $X$ &65 & 0 &   & +1400*  & +1171(30) \\
  $X$ &65 & 2 &   & +1904*$\dag$   & +1825(40)\\
\end{tabular}
\end{ruledtabular}
\end{table}


Ultracold collisions can be modeled with good reliability using the new potentials and the molecular hyperfine and second-order spin-orbit coupling. For $B=0$ the scattering length for the lowest open channel $|1,1\rangle+|1,1\rangle$ is calculated to be $54.54 (20) a_0$ ($a_0=0.529177\times10^{-10}$\,m), which is in a good agreement with $55.1(16) a_0$, derived by Crubellier \textit{et al.} \cite {crubellier1999}, but less well to the value $52.98(40) a_0$ obtained by Samuelis \textit{et al.} \cite{samuelis2000cac} from a spectroscopic study of collision resonances in a molecular beam. Earlier reported values (see e.g.\ Refs.\,\cite{tiesinga1996,abeelen1999doc}) are less precise but agree with the new value within their reported error limits. The small uncertainty for the scattering length results from the improved precision in the Feshbach resonance locations compared to Refs.\,\cite{inouye1998oof,stenger1999sei}. Note that this scattering length is equal to that of the $|1,-1\rangle+|1,-1\rangle$ and $|1,\pm1\rangle+|1,0\rangle$ channels. The other scattering lengths relevant for ultracold Na prepared in the $F=1$ manifold are $52.66(40) a_0$ and $50.78(40) a_0$ for the $|1,0\rangle+|1,0\rangle$ and $|1,1\rangle+|1,-1\rangle$ channels, respectively \cite{scat}.

We can provide a very reliable value for the scattering length of the triplet \astate~state, namely $64.30(40) a_0$. It agrees with the value found in Ref.\,\cite{abeelen1999doc}, but has a 1\,$a_0$ deviation with the value reported in Ref.\,\cite{samuelis2000cac}. This is not surprising given the strong correlation between this scattering lengths and the one for $|1,1\rangle+|1,1\rangle$. The scattering length of the singlet \Xstate~state is mainly determined by the binding energies of the two last bound states $v_X=64$ and 65 \cite{crubellier1999,samuelis2000cac}. Thus we could not improve the precision of the singlet scattering length, but its value has changed slightly to $18.81(80) a_0$. This fact is understandable, because the scattering lengths of the singlet and triplet states are correlated by noting that the $v_X=65$ bound level is significantly mixed with the $v_a=15$ one.

Finally, we propose that the singlet scattering length can be improved by looking for a narrow $p$-wave resonance in collisions between Na atoms in the $|1,1\rangle$ and $|1,0\rangle$ state. We predict a $p$-wave resonance at a magnetic field of 23.5\,mT. It is desirable to improve the molecular beam data from about 30\,MHz to better than 1\,MHz, which now seems feasible given femtosecond comb-based optical frequency measurements.


\begin{acknowledgments}

We acknowledge the workshop, in particular Morris Wei{\ss}er, for the construction of our cloverleaf magnetic trap, Wolfgang Ketterle and Christian Sanner for the loan of an 1.7\,GHz AOM, and Jian Peng Ang and Rico Pires for sharing their ABM code. We acknowledge support from the Heidelberg Center of Quantum Dynamics. E.~Tiemann thanks the
cluster of excellence ``QUEST" for support and the Minister of Science and Culture of Lower Saxony for providing a Niedersachsenprofessur.

\end{acknowledgments}


\begin{thebibliography}{47}
\bibitem[{sca()}]{scat}
\bibinfo{note}{The scattering length values for the coupled basis states
  $|(f_Af_B)f\rangle=|(11)2\rangle$ and $|(11)0\rangle$ are $54.54(20) a_0$ and $48.91(40) a_0$, respectively}.
\end{thebibliography}
\end{document}